\documentclass[prd,twocolumn,nofootinbib,amsmath,amssymb]{revtex4-1}
\usepackage{subfig}
\usepackage{graphicx,bm,hyperref,enumitem,array,dcolumn}
\usepackage{bm}
\usepackage{latexsym}
\usepackage{amssymb}
\usepackage{amsmath}
\usepackage{hyperref}
\usepackage[running]{lineno}

\begin{document} 

\title
{Multigap diffraction cross sections:\\ Problems in eikonal methods for the Pomeron unitarization}

\author{E. Martynov and G. Tersimonov}

\affiliation{%
	Bogolyubov Institute for Theoretical Physics,\\ 03143, Metrologicheskaja st. 14b, Kiev-143, Ukraine.\\
	e-mails: martynov@bitp.kiev.ua;  gters.hep@gmail.com
}%

\date{\today}

\begin{abstract}
{Large rapidity gap diffraction processes are considered in multi-channel eikonal models. It is shown that shadow corrections  to over-fast rising contribution of the input supercritical Pomeron (with $\alpha(0)>1$), originating from the Pomeron rescatterings or, equivalently, accounting survival probability factor, do not solve the Finkelstein-Kajantie problem. Therefore, in our opinion,  another  method of unitarization of supercritical Pomeron should be developed.}  
\end{abstract}
 
\maketitle

\section*{introduction}
Nowadays we are once again witnessing as recent exciting results of the TOTEM experiment on total cross sections of proton-proton interactions at maximal LHC energies \cite{TOTEM} have inspired the big splash of interest in reviving several pending problems unresolved in past \cite{MN, KMR01, CR, BGS}.  One of those is a notorious problem of unitarizing the Pomeron input in various amplitudes of hadronic processes. It has been launched for intensive discussion by studying the BFKL Pomeron \cite{BFKL} having an intercept larger than one in perturbative QCD. Another activity in this scope was waked up by the phenomenological Donnachie-Landshoff model \cite{DL, Eik} that successfully describes the data by the simple Pomeron j-pole located at $t = 0$ above $j = 1$ but badly violates the Froissart-Martin bound \cite{FM}. And finally the recent papers devoted to the multi-Pomeron-odderon vertices (\cite{BCV} and references therein) are seriously focused on the Pomeron unitarization problem. Obviously, the procedure of calculating some corrections for the input Pomeron should be developed to restore an unitarity.
In practical QCD such a calculation, unfortunately, is not possible, and therefore more attention was paid to phenomenological approaches to model a Pomeron in various physical processes. Recently, for example, a special role of Central Exclusive Production (CEP) was intensively investigated because of a possibility to observe the Higgs boson \cite{RKMOR} in such a process, and now \cite{KMR-11} it is investigated to ﬁnd an experimental conﬁrmation of odderon contribution and to study its properties \cite{NKMR}. 

\begin{figure}[!h]
	\centering
	\includegraphics[width=0.9\linewidth]{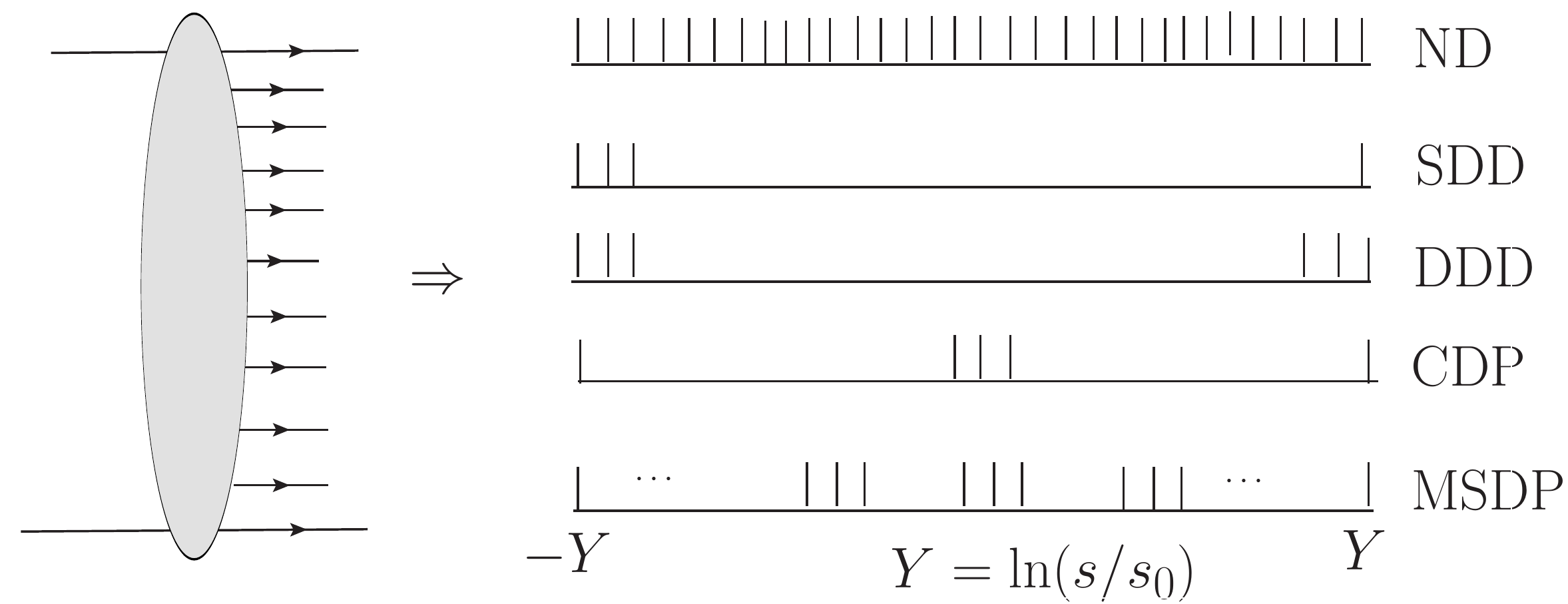}
	\caption{Distribution of the produced hadrons in the various diffraction processes at rapidity scale. ND means NonDiffraction, MSDP means Multi-Showers Diffraction Production, another abbreviations are explained in the text}
	\label{fig:sigd-n-1}
\end{figure}

However, there are a lot of problems with unitarity which are still unresolved even for simpler processes like Simple Diﬀraction Dissociation (SDD), Central Diﬀraction Production (CDP) and Double Diﬀraction Dissociation (DDD) as well as their generalizations including an additional production of high mass showers and large rapidity gaps (LRG) between them. Possible distributions of the produced hadrons are illustrated in the Fig.  \ref{fig:sigd-n-1}. 
 
If the eﬀective masses of produced showers are large enough, then the cross-section of the corresponding processes (together with certain simplifying assumptions) may be be presented (due to the generalized optical theorem) by the diagrams with a triple-Pomeron vertices. The corresponding diagrams are shown in the next Sections and the detailed features of the cross-sections are dependent on speciﬁc model of Pomeron used.
It has been shown long time ago that there is a violation of unitarity bounds for the diﬀraction cross sections even for the ”standard“ simple (in $j$-plane) Pomeron pole with intercept $\alpha_P(0) = 1$ provided that the three-Pomeron vertex is a nonzero constant at zero transfer momentum. It has been formulated as the Finkelstein-Kajantie problem (FK-problem)  \cite{FK,BW} that the contribution of $n$ hadron showers production to the total cross section is increasing with energy as $n$-th power of $\ln(s/s_0)  (s_0 = 1$ GeV) violating the Froissart-Martin bound and self-consistency of the Pomeron model with $\alpha_P(0) = 1$ where $\sigma_{tot}(s) \to const$ at  $s\to \infty$.

One of the most popular and phenomenological successful models at present is the so-called supercritical Pomeron.
In the above mentioned  Donnachie-Landshoff model \cite{DL} Pomeron has a trajectory with the intercept $\alpha_{P}(0)=1+\Delta $ with $\Delta\approx 0.08$. The contribution of such a Pomeron to the total cross-section rises with energy as a power $\sigma \propto s^{\Delta }$, being in a contradiction with the Froissart-Martin bound $\sigma_{tot}<C \ln^{2}(s/s_{0})$. 

The strict and consistent procedure to unitarize Pomeron with an intercept larger than one is unknown until now, but there are some simple phenomenological ways to eliminate the rough contradictions with the unitarity. For example, the eikonal, $U$-matrix methods and their generalizations \cite{Eik,UM,CPS}  are used to input elastic scattering amplitude.

It is quite obvious that any three-Pomeron diagram also needs unitarity corrections, which should remove a too fast-growing contribution of supercritical Pomeron to corresponding diffraction cross-section. The input SDD cross section  is proportional to $s^{2\Delta }$ up to the $\ln s$-factors). The $3P$-diagram seemed to be unitarized by the most simple way, taking into account multiple Pomeron exchanges between the incoming hadrons (initial state interaction). This approach was considered in many old and recent papers Ref. \cite{AK, GLM-1, GLM-2, GLM-3, GKLM, KMR-0, KMR-1, KMR-2, KKMR}. However, we would like to remind here the result of  Ref. \cite{MS}:  the asymptotic estimation of $M^2d\sigma^{SDD}/dtdM^2$ in \cite{GLM-1} is not accurate and has to be corrected.

We will not discuss here other possible approaches to the problem, we concentrate here on the eikonal approach and its modifications. We present here some explicit calculations and high energy estimates within multi-channel eikonal models to check whether the FK-problem is really fixed or not. 

To make our arguments more clear we remind in the Section \ref{sec:elscatt}  some generalities about one-eikonal model and account of rescatterings. We are interested only in an asymptotic cross-section behavior, therefore a contribution of $f$-reggeon is omitted in all expressions. The explicit estimations of corrections to input diffraction cross sections in one-channel eikonal are given in the Section \ref{sec:diffprod}. SDD process within a multi-eikonal model is considered in Section \ref{sec:multieik}.

\section{Elastic Scattering}\label{sec:elscatt}

Following to the Ref. \cite{GLM-1} we will work in the impact parameter representation.  Normalization of elastic scattering amplitude is
\begin{equation}\label{eq:norm}
\dfrac{d\sigma }{dt} =
\pi|F(s,t)|^{2},\qquad \sigma_{tot} = 4\pi \text{Im}F(s,0)
\end{equation}
An amplitude in $b$-representation is defined by the
transformation
\begin{equation}\label{eq:impactampl}
A(s,b) =\dfrac{1}{2\pi }\int d^2{\vec q}e^{-i{\vec q}\,{\vec b}}F(s,t),\quad
t=-q^{2}
\end{equation}
and satisfies the unitarity equation
\begin{equation}\label{eq:unitarity-b}
2\text{Im}A(s,b)=|A(s,b)|^2+G_{inel}(s,b)
\end{equation}
where $G_{inel}(s,b)$ is a contribution of inelastic processes. One can conclude from Eq. \eqref{eq:unitarity-b} that $0<\text{Im}A(s,b)<2$.

Eikonal summation of the high energy elastic Pomeron rescatterings
can be realized with the input amplitude $a(s,b)$ 
\begin{align} \label{eq:eikonalampl}
A(s,b) &= i(1-e^{-\Omega (s,b)}),\\
\Omega (s,b) &=-ia(s,b)=
-\dfrac{i}{2\pi }\int d^2{\vec q}e^{-i{\vec q}\,{\vec b}}f(s,t)
\end{align}
where $f(s,t)$ is an input elastic amplitude.
Starting from a simplified model of supercritical Pomeron
\begin{equation}\label{eq:regge-st-ampl}
f(s,t) = ig^2(t)\left (\dfrac{-is}{s_{0}}\right )^{\alpha(t)-1}\approx ig^2(0)e^{\Delta \xi}e^{(2B_0+\alpha'\xi)t} 
\end{equation}
where 
\begin{equation}\label{eq:pomtrajct}
\xi=\ln(s/s_0), \qquad \alpha(t)=1+\Delta+\alpha't
\end{equation}
and
\begin{equation}\label{eq:vertex}
 g(t)=g(0)\exp(B_0t)
\end{equation}
describes the vertex of Pomeron–proton interaction. 

One can obtain 
\begin{equation}\label{eq:omega}
\Omega (s,b)=\nu (\xi)e^{-b^{2}/R^{2}(\xi)},
\end{equation}
\begin{equation}\label{eq:nu(s)}
\nu (\xi) =\dfrac{2g^2(0)}{R^{2}(\xi)}\biggl(\frac{s}{s_{0}}\biggl)^{\Delta }=
\dfrac{2g^{2}(0)}{R^{2}(\xi)}e^{\Delta \xi},
\end{equation}
\begin{equation}\label{eq:radius2}
R^{2}(\xi) = 4(2B_{0} +\alpha'\ln (s/s_{0}))=4(2B_{0} +\alpha'\xi).
\end{equation}
In this model $if_{0}(s,t)$ and $\Omega (s,b)$ are the real functions. Analyticity and
crossing-symmetry  are restored by the substitution $s\rightarrow
s\exp(-i\pi /2)$. 

It is easy to obtain from the above expressions that at $s\to \infty$
\begin{equation}\label{eq:stoteik}
\begin{aligned}
\sigma_{tot}(s)&=2 \int\limits_0^{\infty}d^2\vec{b}(1-e^{-\Omega(s,b)})\\ &\approx 2\pi \Delta \xi R^{2}(\xi)\to 8\pi \alpha' \Delta \xi^2. 
\end{aligned}
\end{equation}
Thus, in a supercritical Pomeron model the
eikonal corrections to one-Pomeron exchange remove the explicit violation
of unitarity condition for input elastic scattering amplitude \eqref{eq:regge-st-ampl}.

\section{Diffraction production with LRG, one-channel eikonal model}\label{sec:diffprod}
We use for all diffraction cross sections normalization of the Ref. \cite{KTM}.  The difference in the normalization of elastic scattering amplitude here and in \cite{KTM} is taken into account by replacing $g(0)\to g(0)/\sqrt{2}, \quad G_{3P}\to G_{3P}/\sqrt{2}$ in expressions for diffraction cross sections.     

\subsection{Single Diffraction Dissociation}\label{sec:SDD}
\begin{figure}[!hbt]
	\centering
	\includegraphics[width=1.\linewidth]{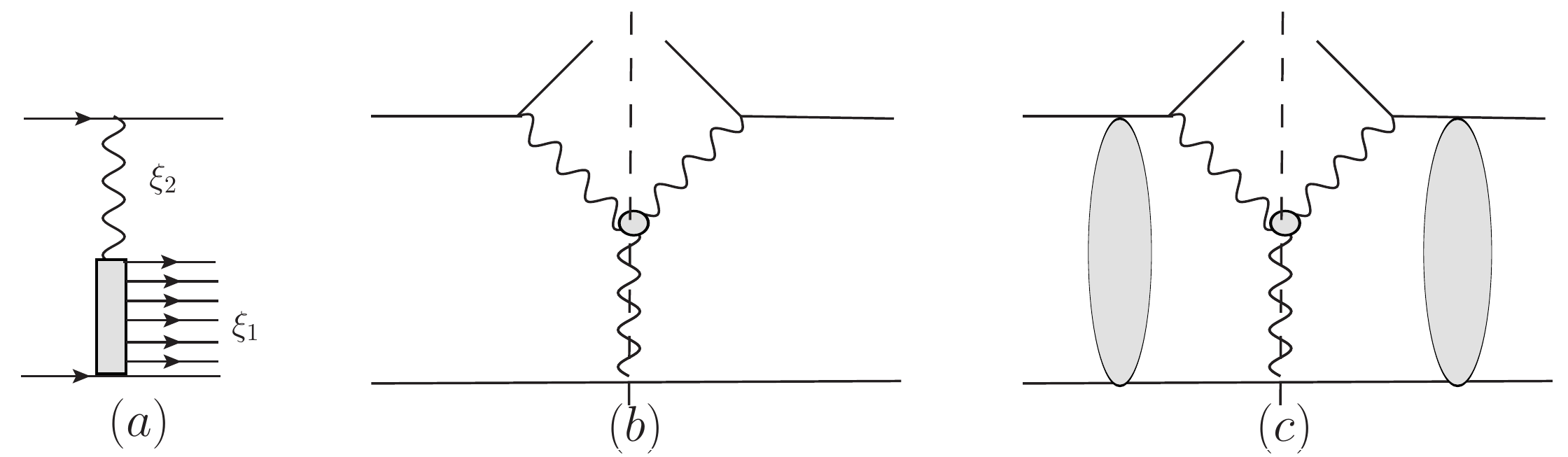}
	\caption{Single Diffraction Dissociation, (a) - SDD process, (b) - 3P diagram for SDD without  survival factor, (c) - 3P diagram with survival factor}
	\label{fig:sdd}
\end{figure}
The input differential SDD cross section pictured in Fig.~\ref{fig:sdd} is written as (let us notice that factorization in $(s,t)-, (s,b)-$ representations is valid for simple $j$-pole)

\begin{equation}\label{eq:3P-0}
\begin{aligned}
M^2\dfrac{d\sigma^{SDD}}{dtdM^2}&=g^2(t)g(0)G_{3P}(0;t,t)\\ &\times \left (\dfrac{s}{M^2}
\right )^{2\alpha_P(t)-2}  \left (\dfrac{M^2}{s_0}
\right )^{\alpha_P(0)-1} 
\end{aligned}
\end{equation}

where 
\begin{equation}\label{eq:3P-vertex}
G_{3P}(t_0;t_1,t_2)=G_{3P}\exp(r^2(t_0+t_1+t_2)
\end{equation} 
is the triple Pomeron vertex.

The expression for an integrated over $t$ cross-section of 
SDD with shadow corrections (or identically, with survival factor) is written in \cite{GLM-1}. With our normalization and notations it is   
\begin{equation}\label{eq:sdd-dcs}
\begin{aligned}
\dfrac{d\sigma^{SDD}}{d\xi_1} & =
2g^{3}(0)G_{3P}
\dfrac{ 2}{\tilde R^{2}(\xi_1)}\left [\dfrac{2}{\tilde R^{2}(\xi_2)}\right ]^{2}e^{\Delta \xi_1+2\Delta \xi_2}\\ 
& \times \displaystyle \int \dfrac{d^2b}{2\pi}\,\dfrac{d^2b'}{2\pi}
\exp\biggl(-2\nu (\xi)e^{-b^{2}/R^{2}(\xi)} \biggr)\\ &\times
\exp\biggl(-\dfrac{({\vec
b}-{\vec{b'}})^{2}}{\tilde R^{2}(\xi_1)}-2\dfrac{b'^{2}}{\tilde
R^{2}(\xi_2)}\biggr)
\end{aligned}
\end{equation}
where $\nu, R^{2}$ are defined by the Eqs.~\eqref{eq:nu(s)},  \eqref{eq:radius2},  $\xi_1=\ln(M^2/s_0)$, $ \xi_2=\ln(s/M^2)$, $\xi_1+\xi_2=\xi=\ln(s/s_0)$,
\begin{equation}\label{eq:R2(s)}
\tilde R^{2}(\xi_i) = 4(B_{0} +r^{2}+ \alpha'\xi_i). 
\end{equation}
The eikonal corrections due to Pomeron rescatterings in initial
state (Fig. \ref{fig:sdd}(b)) were accounted by the insertion of the factor 
\begin{equation}\label{eq:S2(b)}
\exp(-2\Omega (s,b))=\exp\biggl(-2\nu (\xi)e^{-b^{2}/R^{2}(\xi)}\biggr)
\end{equation}
 in the integrand of Eq. \eqref{eq:sdd-dcs}.

It was obtained in \cite{GLM-1} that after integration over $b$ and $b'$ the differential diffraction dissociation cross-section   becomes the following
\begin{equation}\label{eq:sdd-dcs-2}
\dfrac{d\sigma^{SDD}}{d\xi_{1}}=2g^{3}(0) G_{3P}\dfrac{e^{\Delta \xi_1+2\Delta \xi_2}}{\tilde  R^{2}(\xi_2)}a_1\dfrac{\gamma[a_1,2\nu (\xi)]}
{[2\nu (\xi)]^{a_1}}
\end{equation}
where
\begin{equation}\label{eq:sdd-a1}
a_1 = \frac{2R^{2}(\xi)}{2\tilde R^{2}(\xi_1) + \tilde
R^{2}(\xi_2)}
\end{equation}
and $\gamma(a_1,2\nu(\xi))$ is incomplete gamma function, 
$$\gamma(a_1,2\nu(\xi))=\int_0^{2\nu(\xi)} dx x^{a_1-1}e^{-x}.$$
In the limit under consideration,
$s\gg s_{0}$,  $M^{2}/s_{0}, s/M^{2}\gg 1,$ the ratio $a_1$ tends to $2$
and $\gamma[a_1,2\nu(\xi)]$ tends to $\Gamma (2)$. Substituting these limits
to the expression (\ref{eq:sdd-dcs-2}), authors of \cite{GLM-1} had obtained
\begin{equation}\label{eq:eq:sdd-dcs-3}
\dfrac{d\sigma^{SD}}{d\xi_1} \approx \dfrac{G_{3P}R^{4}(\xi)}{4g(0)\tilde R^2(\xi-\xi_1)}e^{-\Delta  \xi_1 }.
\end{equation}
However, this result is wrong. The difference between $a_1$ and 2 is very significant when evaluating the factor $[2\nu(\xi)]^{-a_1}$ in the Eq.~\ref{eq:sdd-dcs-2} in the kinematic region under consideration. Indeed, one can see using the definitions (\ref{eq:regge-st-ampl}),(\ref{eq:omega}),(\ref{eq:radius2}) and (\ref{eq:R2(s)}) that at
$s,\,M^{2},\,s/M^{2} \rightarrow \infty$ (here and in what follows it is sufficient to consider the region where  $\alpha'\xi \gg 2B_0, \quad \alpha'\xi _i\gg B_0+r^2$)
\begin{equation}\label{eq:sdd-a1-2}
a_1 \approx \frac{2\xi}{2\xi_1+ \xi_2} =2 \dfrac{\xi}{\xi+\xi_1}=2\biggl(1 - \frac{\xi_1}{\xi} + \frac{\xi^2_1}{\xi^2}+o\left (\frac{\xi^2_1}{\xi^2} \right )\biggr).
\end{equation}
Therefore the factor in the expression \eqref{eq:sdd-dcs-2}  that violates the unitarity is
transformed as following (we remind here that $\xi_2=\xi-\xi_1$)
\begin{equation}\label{eq:sdd-estmation}
\begin{array}{ll}
\dfrac{e^{\Delta [\xi_1+2\xi_2]}}{[\nu (\xi)]^{a_1}}&\propto
\xi^{a_1}\exp\biggl \{\Delta \left [\xi_1+2\xi_2-
a_1 \xi \right ]  \biggr\}\\
&\approx \xi^2\exp\biggl
\{\Delta \left [\xi_1+2 \xi_2-
2\biggl(1 - \dfrac{\xi_1}{\xi}+ \dfrac{\xi^2_1}{\xi^2}\biggr)\xi \right ]\biggr\}\\
&\\
&=
\xi^2\exp\left \{\Delta\xi_1 [1+O(\xi_1/\xi) ]\right \}\approx \xi^2\exp(\Delta  \xi_1),
\end{array}
\end{equation} 
conserving the fast growth of the SDD cross-section \eqref{eq:sdd-dcs-2} at  $\xi_1=\ln(M^2/s_0)\to \infty$.

\subsubsection{Another definition of survival probability in one-channel eikonal model}\label{sect:SP-one-eik}
This subsection reproduces the part of Ref. \cite{GKLM} concerning the  usage of alternative definition of the survival probability (averaged in $b$) for SDD process. We have  changed only some notations for some variables in Eq.\eqref{eq:Tout-b}.  

\begin{figure}[!hbt]\label{fig:SDD-proc}
	\subfloat[\label{fig:3psimple} SDD without corrections]
	{
		\includegraphics[width=0.5\linewidth]{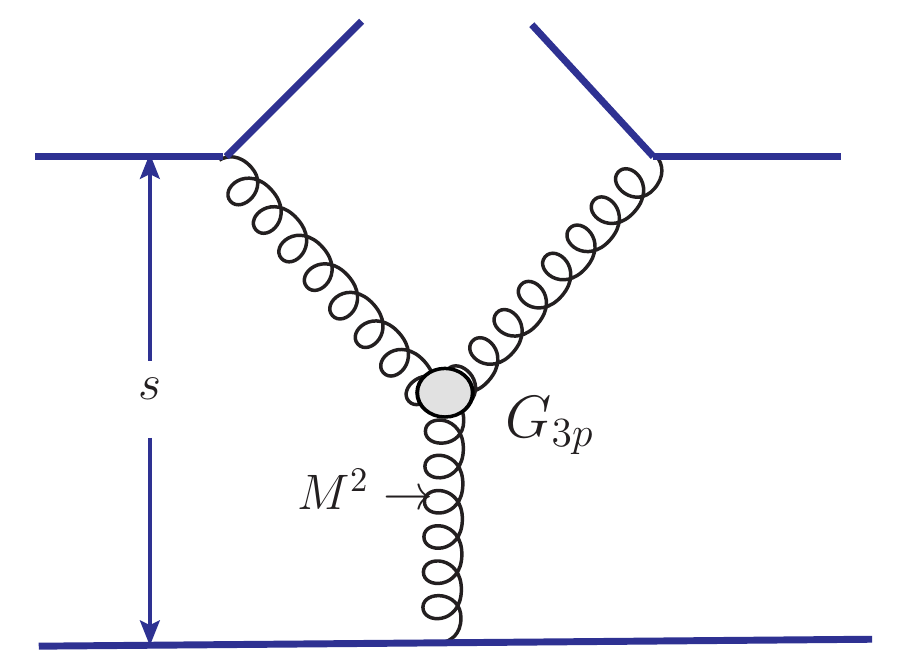}%
	} \\
	\subfloat[\label{fig:3P+SurvP} SDD with survival probability]
	{
		\includegraphics[width=0.5\linewidth]{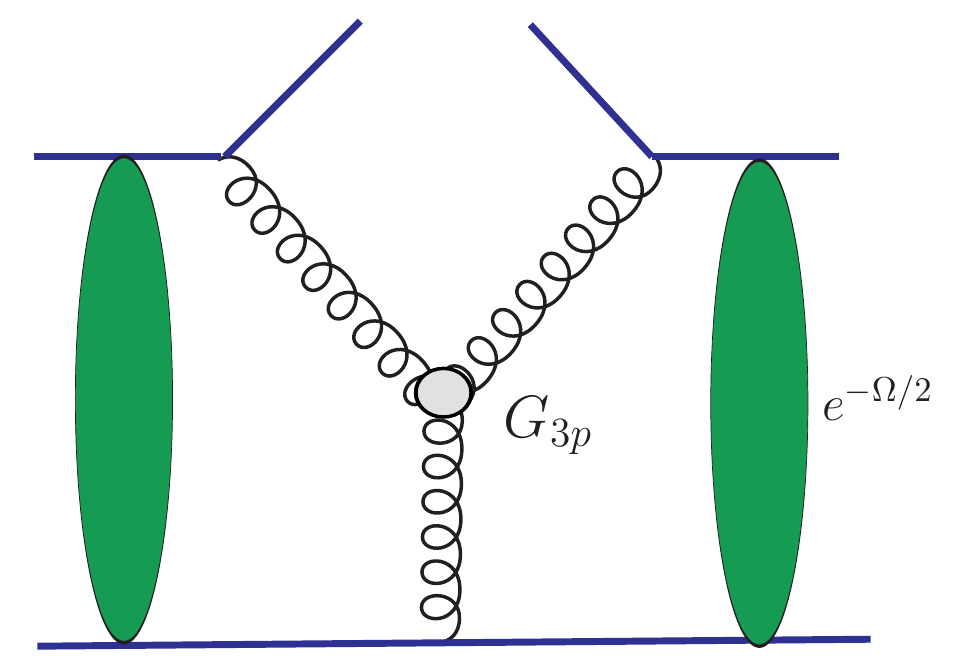}%
	}\\
	\caption{The general Mueller diagram for SDD process in $h–h$ collisions at high energy, (a): without rescatterings and (b): with rescatterings. The spiral lines denote input Pomerons.} 

\end{figure}

The input cross section for diffraction dissociation in the region of large $M^2$ can be viewed as a Mueller diagram (Fig.~\ref{fig:3psimple}) which has been written in Eq. \eqref{eq:3P-0}. We denote the corresponding survival probability at given $M^2$ as $S_{3P}(M^2)$.

The diagram in Fig. \ref{fig:3psimple}. does not take into account a possibility of additional rescatterings of the interacting particles shown in Fig. \ref{fig:3P+SurvP}.  
Contribution of the diagram Fig.~\ref{fig:3psimple} to differential SDD cross section without shadow corrections is given by Eq. \eqref{eq:3P-0}. While the SDD cross section corresponding to Fig.~\ref{fig:3P+SurvP} can be written \cite{GKLM} as 
\begin{equation}\label{eq:3P-1}
\begin{aligned}
M^2\dfrac{d\sigma^{3p}}{dtdM^2}(\text{Fig.} \ref{fig:3P+SurvP})&=S_{3P}^2(M^2)g^2(t)g(0)G_{3P}(0;t,t)\\
&\times \left (\dfrac{s}{M^2}
\right )^{2\alpha_P(t)-2}  \left (\dfrac{M^2}{s_0}
\right )^{\alpha_P(0)-1}
\end{aligned}
\end{equation}
where the  survival probability factor  $S_{3P}^2(M^2)$ averaged over $b$  is defined  as
\begin{equation}\label{eq:SP-def}
S_{3P}^2(M^2)=\dfrac{\int d^2kM^2\dfrac{d\sigma^{3P}}{d^2kdM^2} (\text{Fig.}   \ref{fig:3P+SurvP})}{\int d^2kM^2\dfrac{d\sigma^{3P}}{d^2kdM^2} (\text{Fig.} \ref{fig:3psimple})}, \quad 
t=-k^2.
\end{equation}

The easiest way to calculate the diagram of Fig.~\ref{fig:3P+SurvP} is at first to transform the diagram of Fig.~\ref{fig:3psimple} to impact parameter space. This is done by introducing the momentum $q$ along the lowest Pomeron in Fig.~\ref{fig:3psimple}. In this case 
\begin{equation}\label{eq:Tb}
\begin{array}{ll}
T(s,M^2;q)&=\displaystyle \int d^2k \dfrac{d\sigma^{3P}}{d^2kd\xi_1} (\text{Fig.} \ref{fig:3psimple})
=g^3(0)G_{3P}\\ 
&\times \exp((\xi_1+2\xi_2)\Delta)\exp\left (-q^2\tilde R^2(\xi_1)/4\right)\\ 
&\times \displaystyle \int d^2k \exp\left (-[k^2+(\vec{q}-\vec{k})^2]\tilde R^2(\xi_2)/4\right).
\end{array}
\end{equation}
Similarly to transformation \eqref{eq:impactampl} we find the form of this amplitude in the impact parameter space 
\begin{equation}\label{eq:Tin-b}
F(s,M^2;b)= \int \dfrac{d^2q}{2\pi}e^{-i\vec{q}\vec{b}}A(s.M^2;q)
\end{equation}
Using a linear approximation for the Pomeron trajectory \eqref{eq:pomtrajct} and a Gaussian form for all vertices \eqref{eq:vertex}, \eqref{eq:3P-vertex}
we obtain
\begin{equation}\label{eq:Tout-b}
\begin{array}{ll}
F(s,M^2;b)&=2g^3(0)G_{3P}\dfrac{\nu_0(\xi_1)\nu_0^2(\xi_2)}{d(\xi_1)+2d(\xi_2)}\\
&\times \exp\left (-2\dfrac{d(\xi_1)[d(\xi_2)]}{d(\xi_1)+2d(\xi_2)}b^2\right ),
\end{array}
\end{equation}
where
\begin{equation}\label{eq:defvars-c}
\nu_0(y)=\dfrac{2e^{\Delta y}}{\tilde R^2(y)},\quad
d(y)\equiv \dfrac{1}{\bar R^2(y)}.
\end{equation}

Making use of the Eq. \eqref{eq:Tout-b} the expression for the survival probability  \eqref{eq:SP-def} in a simple eikonal model with the rescattering corrections can be written as
\begin{equation}\label{eq:SP-b}
S_{3P}^2(M^2)=\dfrac{\int d^2b F(s,M^2;b)\exp(-\Omega(\xi;b))}{\int d^2b F(s,M^2;b)} \end{equation}
where
\begin{equation}
\Omega(\xi;b)=2\nu(\xi)\exp\left (-\dfrac{b^2}{R^2(\xi)} \right ), \quad
\nu(\xi)=\dfrac{2g^2(0)}{R^2(\xi)}e^{\Delta \xi}. 
\end{equation}

{\bf Caculation of $S_{3P}^2(M^2)$ and $M^2\dfrac{d\sigma^{3p}}{dtdM^2}$}

The above defined averaged survival probablityt is presented (with some nonprincipal modifications) in many papers as the unitarization method  (or the compensation of  the too fast increasing with energy cross section) for the Pomeron with $\alpha_{P}((0)-1=\Delta>0$. Let's check if this procedure indeed compensates too fast growth and really solves the Finkelstein-Kajantie problem. 

The averaged survival probability $S_{3P}^2(M^2)$ \eqref{eq:SP-b}   is easily  calculated  and estimated  in the limit under interest. 
\begin{equation}\label{eq:SP-M2}
S_{3P}^2(M^2)\approx \dfrac{e^{\Delta \xi_1+2\Delta \xi_2}}{\tilde  R^{2}(\xi_2)}a_1\dfrac{\gamma[a_1,2\nu (\xi)]}
{[2\nu (\xi)]^{a_1}}\Big /\frac{ e^{\Delta \xi_1 }
	e^{2\Delta \xi_2}}{\tilde R^{2}(\xi_1)[\tilde R^{2}(\xi_2)]^{2}}
\end{equation}
where $a_1$ and its asymptotic  are determined by Eqs, \eqref{eq:sdd-a1} and \eqref{eq:sdd-a1-2}.  Omitting the constant and logarithmic factors we obtain
\begin{equation}\label{eq:SP-M2-as}
\begin{aligned}
S_{3P}^2(M^2)&\propto (\nu(\xi))^{a_1}\approx \exp(-2(1-\xi_1/\xi)\Delta \xi)\\&=\exp(-2\Delta \xi_2).
\end{aligned}
\end{equation}
Then, it  follows  from \eqref{eq:3P-1} that
\begin{equation}\label{eq:sdd-final}
\dfrac{d\sigma^{3p}}{d\xi_1}\propto\exp(\Delta (\xi_1+2\xi_2))\exp(-2\Delta \xi_2)=\exp(\Delta\xi_1)
\end{equation}
 Conclusion from the Section \ref{sec:SDD}.
An over unitarity growth of  the input SDD cross section is not compensated by  survival probability factor.

\subsection{Central Diffraction Production, CDP}\label{sec:CDP}

\begin{figure}[!hbt]
	\centering
	\includegraphics[width=1.\linewidth]{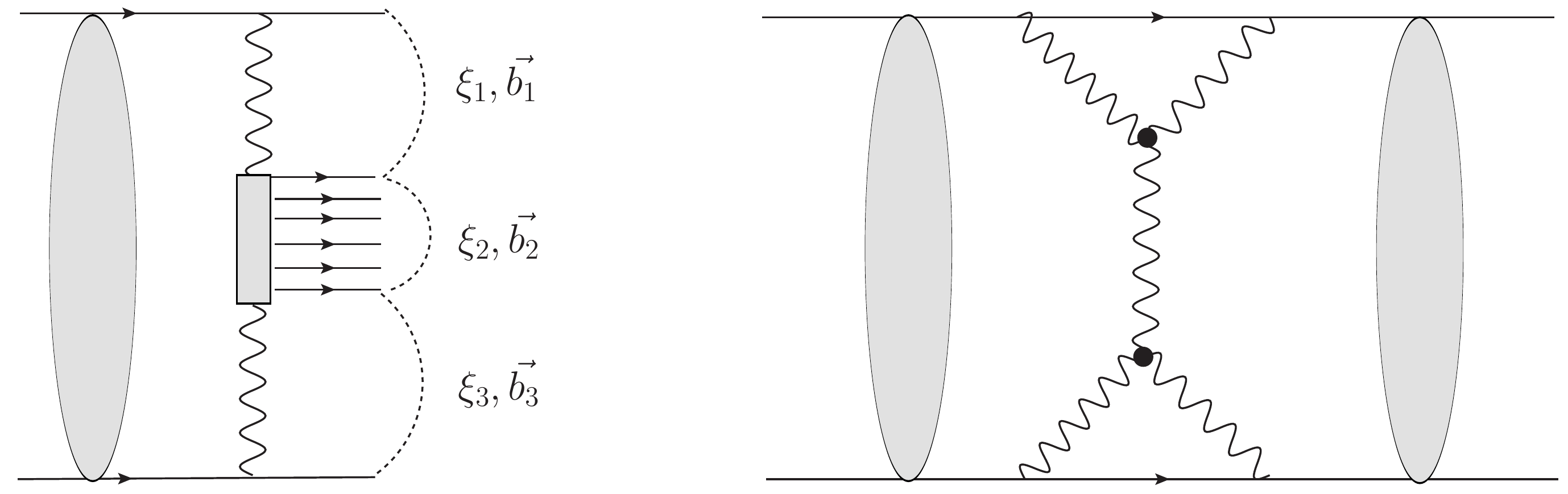}
	\caption[]{Central Diffraction Production}
	\label{fig:cdp}
\end{figure}

Let's consider a process of  the central diffraction  production shown in Fig. \ref{fig:cdp}. 
\begin{equation}\label{eq:cdp-0}
\begin{aligned}
\dfrac{d\sigma^{CDP}}{dt_1dt_2d\xi_2}&=\dfrac{1}{4\pi}g^4(0)G^2_{3P}e^{\Delta (2\xi_1+\xi_2+2\xi_3)}\\ &\times \exp(-q_1^2\tilde R^2(\xi_1)/2)\exp(-q_2^2\tilde R^2(\xi_3)/2)
\end{aligned}
\end{equation}

The differential CDP cross  section  integrated over  $t_1$ and  $t_2$  is written in terms of the impact parameters as follows
\begin{equation}\label{eq:cpd-dcs}
\begin{array}{ll}
\dfrac{d\sigma^{CDP}}{d\xi_1d\xi_2}&=
4\pi g^4(0)G^2_{3P}
\left [\dfrac{2e^{\Delta \xi_1}} {\tilde R^{2}(\xi_1)}\right ]^{2}\dfrac{e^{\Delta \xi_2}} {\tilde R^{2}(\xi_2)}\left [\dfrac{2e^{\Delta \xi_3}}{\tilde R^{2}(\xi_3)}\right ]^{2} \\
& \times \displaystyle \int \dfrac{d^2b}{2\pi}\,\dfrac{d^2b_1}{2\pi}\,\dfrac{d^2b}{2\pi}
\exp\biggl(-2\nu (\xi)e^{-b^{2}/R^{2}(\xi)}\biggr)\\
&\times \exp\biggl (-2\dfrac{\vec{b_1}^2}{\tilde R^{2}(\xi_1)}-\dfrac{\vec{b_2}^2}{\tilde
	R^2(\xi_2)}-2\dfrac{\vec {b_3}^2}{\tilde R^2(\xi_3)} \biggr)
\end{array}
\end{equation}
where $\xi_2=\xi-\xi_1-\xi_3, \quad \vec{b_2}=\vec{b}-\vec{b_1}-\vec{b_3}$.

Performing the integration in  the Eq. (\ref{eq:cpd-dcs}) one  can obtain
\begin{equation}\label{eq:cpd-dcs-2}
\begin{aligned}
\dfrac{d\sigma^{CDP}}{d\xi_1d\xi_2}&=16\pi g^4(0)G^2_{3P}\dfrac{R^2(\xi)}{\tilde R^2(\xi_1)\tilde R^2(\xi_3)}a_2\\ &\times [2\nu (\xi)]^{-a_2}\gamma(a_2,2\nu(\xi)) e^{2\Delta \xi_1 }
e^{\Delta \xi_2}e^{2\Delta \xi_3 }
\end{aligned}
\end{equation} 
where
\begin{equation}\label{eq:a-2}
a_2=\frac{2R^{2}(\xi)}{\tilde R^{2}(\xi_1) +2 \tilde R^{2}(\xi_2) + \tilde R^{2}(\xi_3)}
\end{equation} 
Like to SDD case $a_2\to 2$ at $\xi\to \infty$, however taking into account that $R^2(\xi_i)\approx  \alpha'\xi_i$ at $\xi \gg 1$ we have
\begin{equation}\label{eq:a-2estimation}
a_2\approx 2\dfrac{\xi}{\xi_1 + 2\xi_2+ \xi_3}=2\dfrac{1}{1+\xi_2/\xi}\approx  2(1-\xi_2/\xi). 
\end{equation} 

As a result we see that the corrected CPD cross section
\begin{equation}\label{eq:cpd-dcs-3}
\begin{aligned}
\dfrac{d\sigma^{CDP}}{d\xi_1d\xi_2}&\propto (\nu(\xi))^{-a_2}e^{2\Delta \xi_1 }e^{\Delta \xi_2}e^{2\Delta \xi_3 }\\ &\approx e^{[\Delta (2\xi-\xi_2-2(\xi-\xi_2))]} =\exp(\Delta\xi_2)
\end{aligned}
\end{equation}  
rises faster than it is allowed by unitarity.

Conclusion from the Section \ref{sec:CDP}: An over unitarity growth of  the input CDP cross section is not compensated by  survival probability factor.

\subsection{Double Diffraction Dissociation, DDD}\label{sec:DDD}

\begin{figure}[!hbt]
	\centering
	\includegraphics[width=0.9\linewidth]{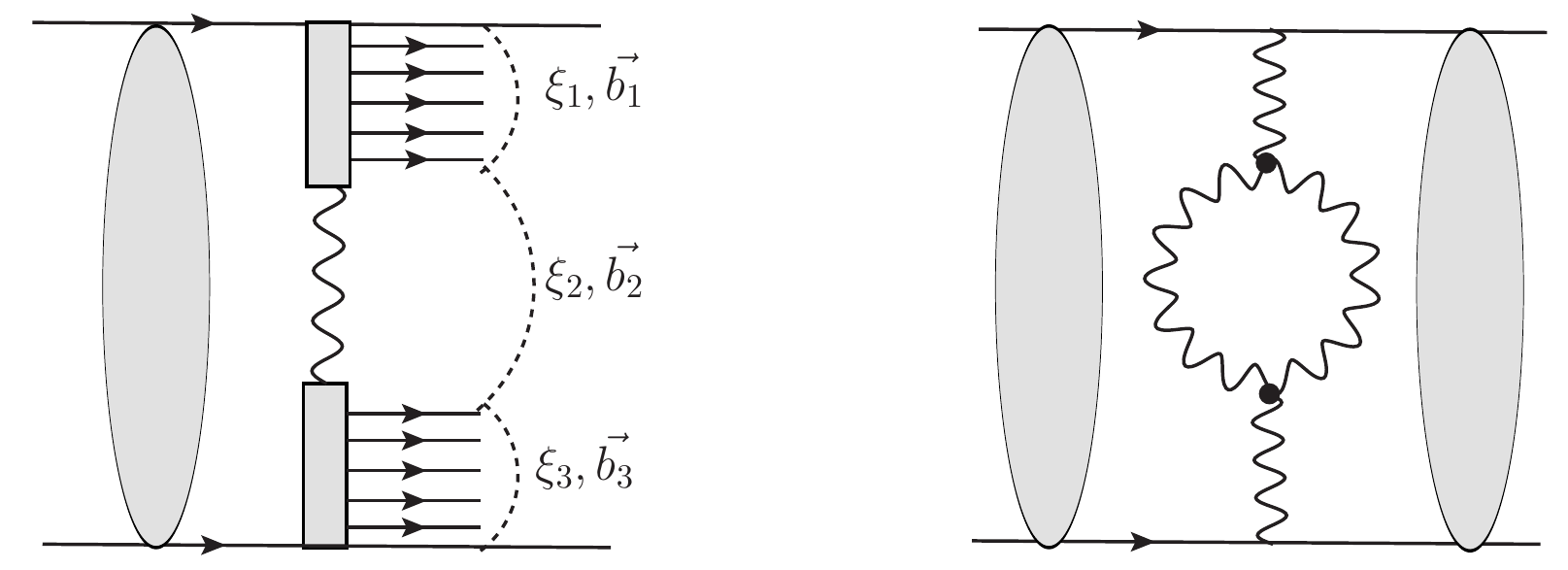}
	\caption[]{Double Diffraction Dissociation}
	\label{fig:dd}
\end{figure}

DDD cross section without rescatterings is  calculated by the following expression
\begin{equation}\label{eq:ddd-0}
\dfrac{d\sigma^{DDD}}{d\xi_1d\xi_3dt}=g^2(0)G^2_{3P}e^{\Delta (\xi_1+2\xi_2+\xi_3)}\exp(-2q^2\tilde R^2(\xi_2))
\end{equation}
The integrated over $t$ DDD cross section with rescatterings has the form 
\begin{equation}\label{eq:ddd-dcs}
\begin{array}{ll}
\dfrac{d\sigma^{DDD}}{d\xi_1d\xi_2}  &=
g^2G^2_{3P}
\dfrac{2e^{\Delta \xi_1}}{\tilde R^{2}(\xi_1)}\left [\dfrac{2e^{\Delta \xi_2}}{\tilde R^{2}(\xi_2)}\right]^2\dfrac{2e^{\Delta \xi_3}}{ \tilde R^{2}(\xi_3)}\\ 
& \times \displaystyle \int  \dfrac{d^2b}{2\pi}\,\dfrac{d^2 b_1}{2\pi}\,\dfrac{d^2b_2}{2\pi}
\exp\biggl(-2\nu (\xi)e^{-b^{2}/R^{2}(\xi)}\biggr)\\
&\times
\exp\biggl (-\dfrac{\vec{b_1}^2}{\tilde R^{2}(\xi_1)}-2\dfrac{\vec{b_2}^2}{\tilde
	R^2(\xi_2)}-\dfrac{\vec {b_3}^2}{\tilde R^2(\xi_3)} \biggr)
\end{array}
\end{equation}
Similarly to the previous calculations we obtain
\begin{equation}\label{eq:ddd-dcs-2}
\begin{aligned}
\dfrac{d\sigma^{DDD}}{d\xi_1d\xi_2} &=2g^2G^2_{3P}\dfrac{R^2(\xi)}{\tilde R^2(\xi_2)}a_3\\ &\times[2\nu (\xi)]^{-a_2}\gamma(a_3,2\nu(\xi)) e^{\Delta (\xi_1
+2\xi_2+\xi_3)}
\end{aligned}
\end{equation}
\begin{equation}\label{eq:a-3}
\begin{aligned}
a_3&=\frac{2R^{2}(\xi)}{2\tilde R^{2}(\xi_1) + \tilde R^{2}(\xi_2) + 2\tilde R^{2}(\xi_3)}\\ &\approx \dfrac{2}{2\xi_1/\xi + \xi_2/\xi +2 \xi_3/\xi}=\dfrac{2}{1+\xi_1/\xi+\xi_3/\xi}\\   \nonumber
&\approx  2(1-\xi_1/\xi-\xi_3/\xi) \nonumber
\end{aligned}
\end{equation}  
\begin{equation}\label{eq:ddd-dcs-3}
\begin{aligned}
\dfrac{d\sigma^{DDD}}{d\xi_1d\xi_3}&\propto (\nu(\xi))^{-a_3}e^{\Delta (\xi_1+\xi_2 + \xi_3)}\\ &\approx e^{\Delta [-2(\xi-\xi_1-\xi_3)+2\xi+\xi_1+2\xi_2+\xi_3]}\\  \nonumber
&=\exp(\Delta(\xi_1+\xi_3)).
\end{aligned}
\end{equation}
Again we  have a violation of unitarity. There is no compensation  of  too fast rising  input contribution of the Pomeron with intercept $\alpha(0)=1+\Delta>1$.

Conclusion from the Section \ref{sec:DDD}. An over unitarity growth of  the input DDD cross section is not compensated by  survival probability factor.
.

\subsubsection{Double Diffraction Dissociation, with additional many LRG showers}\label{sect:DDD-n}

\begin{figure}[!hbt]
	\centering
	\includegraphics[width=0.6\linewidth]{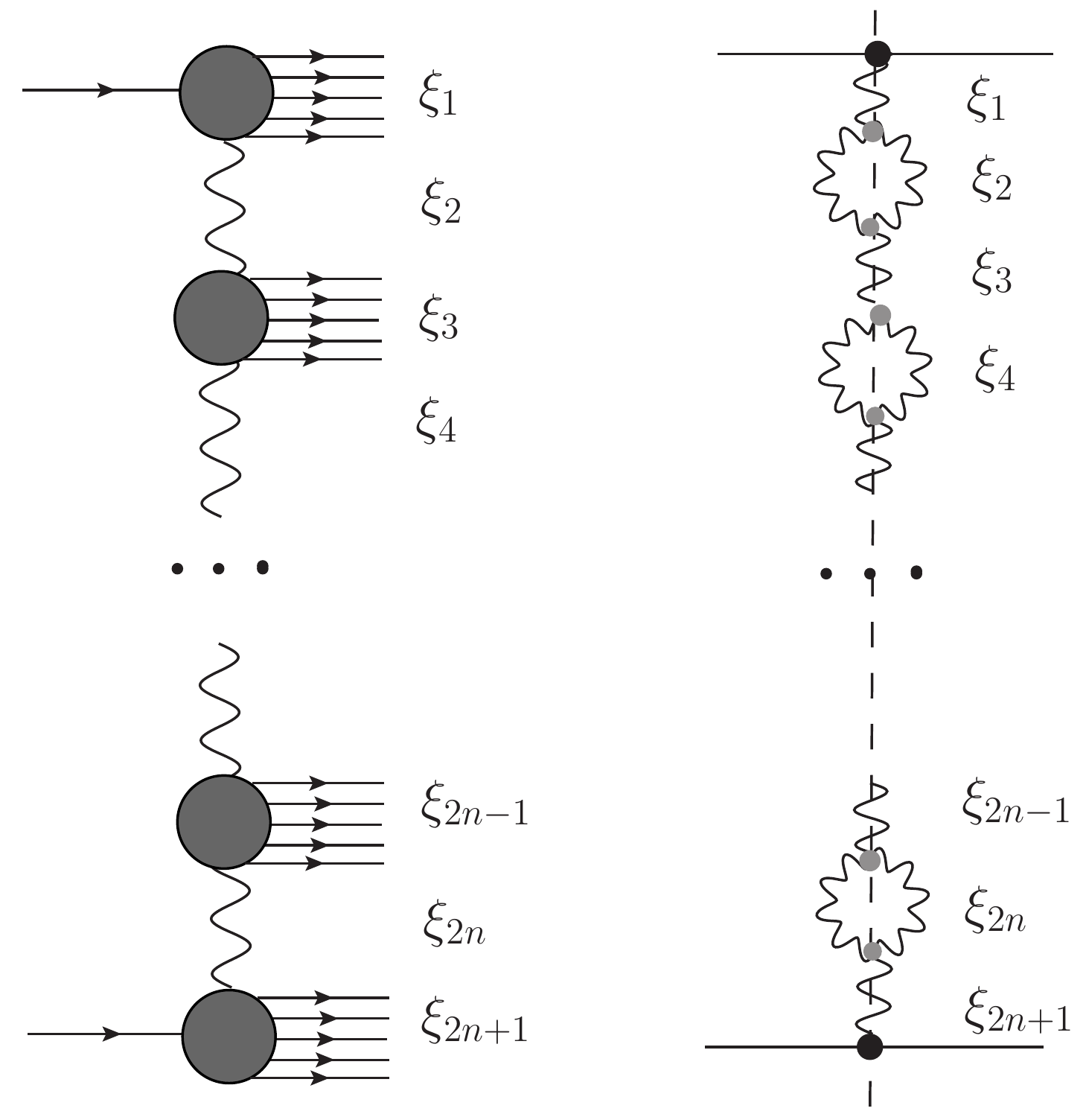}
	\caption{Process of diffraction $n$-showers production}
	\label{fig:dd-n}
\end{figure}

We write differential DDDn cross section (Fig.~ \ref{fig:dd-n}) as follows
\begin{equation}\label{eq:ddd-n-t}
\begin{aligned}
\dfrac{d\sigma^{DDDn}}{d\xi_1\cdots d\xi_{2n}dt_1\cdots dt_n}&=4\pi g^2(0)\left ( \dfrac{G^2_{3P}}{4\pi}\right )^{n}\\
&\times \exp\left (\Delta \sum\limits_{i=0}^n \xi_{2i+1}+2
\Delta \sum\limits_{i=1}^n\xi_{2i}\right )\\&\times \prod\limits_{i=1}^{n}\exp\left (-q_i^2\tilde R^2(\xi_{2i})/2\right ).
\end{aligned}
\end{equation}
This cross section integrated over $t_i$ in $b$-representation:
\begin{equation}\label{eq:ddd-n}
\begin{aligned}
\dfrac{d\sigma^{DDDn}}{d\xi_1d\xi_2\cdots d\xi_{2n}}&=
4\pi g^2(0)\left ( \dfrac{G^2_{3P}}{4\pi}\right )^{n}\\&\times 
\prod\limits_{i=1}^{n+1}\dfrac{2e^{\Delta\xi_{2i-1}}}{\tilde R^{2}(\xi_{2i-1})}\prod\limits_{i=1}^n\left [\dfrac{2e^{\Delta\xi_{2i}}}{[ \tilde R^{2}(\xi_{2i})}\right ]^2\\
&\times \displaystyle \int \dfrac{d^2b}{2\pi}\,\dfrac{d^2b_1}{2\pi}\cdots \dfrac{d^2b_{2n+1}}{2\pi}\delta\left ({\vec b}-\sum\limits_{i=1}^{2n+1}{\vec b_i}\right )\\
&\times(2\pi)^2 \exp\biggl(-2\nu (\xi)e^{-b^{2}/R^{2}(\xi)}\biggr)\\&\times 
\exp\biggl (-\sum\limits_{i=1}^{n+1}\dfrac{\vec{b}^2_{2i-1}}{\tilde R^{2}(\xi_{2i-1})}-2\sum\limits_{i=1}^n\dfrac{\vec{b}^2_{2i}}{\tilde
	R^2(\xi_{2i})}\biggr ) 
\end{aligned}
\end{equation} 

After integration over all $\vec{b}$-s we have 
\begin{equation}\label{eq:ddd-n-2}
\begin{aligned}
\dfrac{d\sigma^{DDDn}}{d\xi_1d\xi_2\cdots d\xi_{2n}}& =8\pi g^2(0)\left ( \dfrac{G^2_{3P}}{2\pi}\right )^{n}\\ &\times R^2(\xi)\exp\left (\Delta \left (\sum\limits_{i=0}^n \xi_{2i+1}+2
\sum\limits_{i=1}^n\xi_{2i}\right )\right )\\
&\times \dfrac{e^{2\Delta\xi_{2i}}}{\tilde R^{2}(\xi_{2i})}a_n[2\nu (\xi)]^{-a_n}\gamma(a_n,2\nu(\xi)) 
\end{aligned}
\end{equation}
\begin{equation}\label{eq:a-n}
\begin{aligned}
a_n\approx 2\dfrac{1}{2\sum\limits_{i=1}^{n+1} \xi_{2i-1}/\xi + \sum\limits_{i=1}^{n} \xi_{2i}/\xi }\approx  2\left (1-\sum\limits_{i=1}^{n+1} \xi_{2i-1}/\xi \right )
\end{aligned}
\end{equation}
\begin{equation}\label{eq:ddd-n-3}
\begin{aligned}
\dfrac{d\sigma^{DDDn}}{d\xi_1d\xi_2\cdots d\xi_{2n}}&\propto (\nu(\xi))^{-a_n}\\&\times \exp\left (\Delta \left (\sum\limits_{i=1}^{n+1} \xi_{2i-1} + 2\sum\limits_{i=1}^{n}\xi_{2i}\right )\right )\\
&\approx \exp\left(\Delta \sum\limits_{i=1}^{n+1} \xi_{2i-1}\right ).
\end{aligned}
\end{equation}

So, the one-eikonal ''survival probability method`` of unitarity restoration does not work for multi-shower generalization of DDD process.  The similar conclusion can be obtained for SDDn and CDPn processes. 

The general conclusion of the Section \ref{sec:diffprod}. We have argued that the FK problem for the main diffraction processes is not fixed by one-channel eikonal survival probability unitarization.
\section{Diffraction production with LRG in two-channel eikonal model}\label{sec:multieik}
In this Section we consider an unitarization of the SDD cross section in the framework of two-channel eikonal model following the paper \cite{KMR-0} (similar model is considered in \cite{GKLM}). 

Let us briefly remind  the main idea of the method following to the Ref. \cite{KMR-0}.

Authors have used a two-channel eikonal (see also \cite{AK}) in which, besides the elastic proton channel  proton excitation $N^*$, a possible intermediate state in $pp$ elastic scattering, is allowed. This {\it effective} $N^*$ channel describes the sum of low mass diffractive proton excitations.  For the various $p$ and $N^*$ couplings to the  Pomeron a common dependence on $t$ is taken    
\begin{equation}\label{eq:append1}   
\begin{aligned}  
\beta_p\rightarrow \left (\begin{array}{ll}     
	\beta (p \rightarrow p) & \beta (p \rightarrow N^*) \\    
	\beta (N^*\rightarrow p) & \beta (N^*\rightarrow N^*) \end{array}\right ) \simeq \beta     
(p\rightarrow p)\left (\begin{array}{cc} 1&\gamma \\ \gamma &1 \end{array} \right )    
\end{aligned}
\end{equation}    
where 
\begin{equation}    
\label{eq:a12}    
\gamma \equiv \frac{V(p\rightarrow N^*)}{V(p\rightarrow p)},    
\end{equation}    

Here for asymptotic estimates the simplest choice for the vertex can be used $\beta(t)=\beta_p\exp(B_0t)$  and Pomeron trajectory $\alpha(t)=1+\Delta+\alpha't$. 

Now each amplitude has two vertices and so, for the amplitudes under consideration we have
\begin{equation} \label{eq:append4}   
 \begin{aligned}   
{\rm Im} A_{\rm el} (b) &=1-\frac{1}{4} \left [ e^{- (1+\gamma)^2     
	\Omega}+2 e^{-(1-\gamma^2)\Omega}\right .\\&\left . +e^{-(1-\gamma)^2\Omega}\right],\\    
{\rm Im} A (pp \rightarrow N^* p) &=\frac{1}{4} \left [ e^{-(1-\gamma)^2	\Omega}-e^{-(1+\gamma)^2 \Omega} \right ],\\     
{\rm Im} A (pp \rightarrow N^* N^*)&= \frac{1}{4} \left [ e^{- (1 - \gamma)^2     
	\Omega} -2 \: e^{- (1 - \gamma^2) \Omega}\right .\\&\left .+e^{- (1 + \gamma)^2    
	\Omega} \right ]. 
\end{aligned}
\end{equation}    
 $\Omega \equiv \Omega (s, b)$ is defined by Eqs. \eqref{eq:omega}, \eqref{eq:nu(s)}, \eqref{eq:radius2}.
\begin{equation}\label{eq:sdd-SP}
\begin{aligned}
\dfrac{d\sigma^{SDD}_E}{d\xi_1}&=16 g^{3}_p(0)G_{3P}
\dfrac{ e^{2\Delta \xi_2 }
	e^{\Delta \xi_1}}{\tilde R^{2}(\xi_1)[\tilde R^{2}(\xi_2)]^{2} }\displaystyle \int \dfrac{d^2b}{2\pi}\,\dfrac{d^2b'}{2\pi}\\ 
&\times 
E(\Omega)
\exp\biggl(-\dfrac{({\vec
		b}-{\vec{b'}})^{2}}{\tilde R^{2}(\xi_1)}-2\dfrac{b'^{2}}{\tilde
	R^{2}(\xi_2)}\biggr)
\end{aligned}
\end{equation}
In the considered two-channel eikonal model \cite{KMR-0}
\begin{equation} \label{eq:append19}   
\begin{aligned}    
E(\Omega)=&\frac{1}{8} \left \{ (1+\gamma) \left [(1+ \gamma)e^{-(1+ \gamma)^2 \Omega/2}\right . \right . \\ &\left . \left . +(1 -\gamma) e^{-(1-\gamma^2) \Omega/2} \right ]^2    
\right .\\    
&+(1-\gamma) \left . \left [ (1-\gamma)e^{-(1-\gamma)^2 \Omega/2}\right . \right .\\&\left . \left .  +    
(1+\gamma) e^{-(1 -\gamma^2) \Omega/2} \right ]^2 \right \}.\    
\end{aligned}
\end{equation}    
Obviously, $E(\Omega$) is the sum of similar type terms, that can be written in the form $P_{\gamma}e^{-2p_{\gamma}\Omega}$. 
Now we can calculate and estimate asymptotic behavior of any term in the differential cross section of SDD \eqref{eq:sdd-SP}.
\begin{equation}\label{eq:sdd-SPa}
\begin{aligned}
\dfrac{d\sigma^{SDD}_{part}}{d\xi_1}&=
16 g^{3}_p(0)G_{3P}
\dfrac{ e^{2\Delta \xi_1 }
	e^{\Delta \xi_2}}{ \tilde R^{2}(\xi_1)[\tilde R^{2}(\xi_2)]^{2}}P_\gamma\\ 
& \times \displaystyle \int \dfrac{d^2b}{2\pi}\,\dfrac{d^2b'}{2\pi}
e^{-2p_\gamma\nu (\xi)e^{-b^{2}/R^{2}(\xi)}}\\&\times 
\exp\biggl(-\dfrac{({\vec
		b}-{\vec{b'}})^{2}}{\tilde R^{1}(\xi_1)}-2\dfrac{b'^{2}}{\tilde
	R^{2}(\xi_2)}\biggr)
\end{aligned}
\end{equation}
Let us note that the Eq. \eqref{eq:sdd-SPa} almost coincides with Eq. \eqref{eq:sdd-dcs}. Therefore, we have in the two-channel eikonal model for any term of SDD cross section
\begin{equation}\label{eq:sdd-SPb}
\begin{aligned}
\dfrac{d\sigma^{SDD}_{part}}{d\xi_{1}}&=2g^{3}_p(0) G_{3P}P_\gamma\dfrac{e^{\Delta \xi_1+2\Delta \xi_2}}{\tilde  R^{2}(\xi_2)}a_1\\&\times \dfrac{\gamma[a_1,2p_\gamma\nu (\xi)]}
{[2p_\gamma\nu (\xi)]^{a_1}}\propto \xi^2e^{\Delta\xi_1}
\end{aligned}
\end{equation}
where $a_1$ is determined by Eq. \eqref{eq:sdd-a1}. SDD cross section \eqref{eq:sdd-SPb} 
rises as $(M^2/s_0)^\Delta$ and violates the unitarity bound at asymptotic energy.
This result confirms the conclusions made in the previous Section. 

\medskip

\section*{concliusion}
It has been declared in the papers \cite{GLM-1,GLM-2,KMR-1,KMR-2} that the too fast (like power of energy, if $\alpha(0) > 1$) growth of multi-gap diﬀraction production cross section can be compensated within the eikonal approach by including shadow corrections to the amplitude (or the Pomeron rescatterings in initial state), in other words, due to survival probability factor.  It is important that  the considered eikonal models realize the BDL when $\text {Im}A(s,b \approx 0) \to 1$ at $s\to \infty$. 

If it is so, then well-known Finkelstein-Kajantie problem (multi-gap diffraction cross sections rise with energy beyond the unitarity bound) is resolved. We would like to note, that in all eikonal models considered in the cited papers, the final dependence of diffraction cross-sections on the effective mass of produced showers actually was not calculated except perhaps the Ref. \cite{GLM-1} where SDD cross section was estimated, however, far from sufficient accuracy as we demonstrated in the Section \ref{sec:diffprod}.

In fact, we have argued here more accurate estimates of corrections show the opposite trend for the FK compensation. Not only the main eikonalized diﬀraction cross sections (SDD, CDP, DDD) violate unitarity bounds. The eikonalized cross sections of generalized processes with additional production of any number of hadron heavy showers with LRG between them are running into the same failure. Moreover, we have considered two approaches for survival probability factor and neither one-channel eikonal model, nor two-channel model have showed the same, negative, answer as to the FK problem for diﬀraction cross sections. One can see that too fast growth of the cross sections is retained in three-channel eikonal approach \cite{GLM-3}].

Thus, we conclude that the Finkelstein-Kajantie problem is not solved due to survival probability factor within the BDL eikonal approach. 

In our opinion another approach should be developed for unitarization of input supercritical Pomeron in multi-gap diﬀraction processes. Moreover, probably, alternative approach, beyond the eikonal one, should be considered in order to describe multi-gap diﬀraction processes in case of rising total cross section. This approach will be presented in our forthcoming paper.

\acknowledgments {We thank Professors G.~Zinovjev, B.~Nicolescu and S.M.~Troshin for a careful reading of the manuscript and interesting, fruitful discussions. Research funded by National Academy of Sciences of Ukraine (Project No.~ 0120U100935).}

\end{document}